\documentclass[aps,pre,twocolumn,superscriptaddress,showpacs]{revtex4-2}

\usepackage{amsmath,amssymb,amsthm}
\usepackage{graphicx}
\usepackage{hyperref}
\usepackage{booktabs}
\usepackage{xcolor}
\usepackage{bm}
\usepackage{dcolumn}

\newcommand{\avg}[1]{\langle #1 \rangle}
\newcommand{\pc}{p_c}
\newcommand{\DF}{\Delta F}
\newcommand{\kcut}{k_{\text{cut}}}
\newcommand{\keff}{K_{\text{eff}}}

\begin{document}

\title{Compounding Vulnerability: Simultaneous Degradation of Percolation and Cascade Robustness Under Targeted Hub Removal}

\author{Federico Cachero}
\affiliation{Independent Researcher, Fort Lauderdale, FL, USA}
\email{federicohernancachero@gmail.com}

\date{\today}

\begin{abstract}
Targeted hub removal is known to weaken connectivity in heterogeneous networks. We show that in Barab\'asi--Albert networks the same intervention can also shift Watts threshold dynamics across the cascade critical point. For BA networks with $N=2{,}000$ and $m=2$, removing the top 10\% of nodes by degree raises the bond-percolation threshold from $\pc=0.174$ to $0.776$ and, at $\varphi=0.22$, increases mean cascade size from $0.86\%$ (95\% CI 0.43--1.30) to $23.1\%$ (21.3--24.9). A controlled hub-vulnerability experiment on fixed topology shows that most of this cascade effect is dynamical: lowering hub activation thresholds produces much larger cascades even without deleting nodes, while deletion partly offsets the increase by removing edges. Using a configuration-model approximation, we derive the post-removal branching factor $z_1$ and identify a window in which the original network is subcritical but the hub-removed network is supercritical. The effect persists across system sizes and is not seen in matched ER or WS controls. These results identify a regime in which hub removal simultaneously worsens connectivity and cascade exposure in BA networks.
\end{abstract}


\maketitle

\section{Introduction}
\label{sec:intro}

Network resilience---the ability of a system to maintain function under stress---is a central problem of complexity science~\cite{Boccaletti2006,Gao2016,Barabasi2016,Artime2024,Valdez2020}. Resilience is not a property of a graph $G$ alone, but of the pair $(G, S)$ consisting of a network and a specified stress model~\cite{Alderson2010}. Two canonical stress models illustrate this~\cite{Stauffer2014,Newman2010}: (i) bond percolation, where edges are retained with probability $p$ and the observable is the giant component size, yielding the critical threshold $\pc$ as a fragility metric; and (ii) the Watts threshold cascade~\cite{Watts2002}, where nodes activate when a fraction $\varphi$ of their neighbors are active, and the observable is the final cascade size.

Two research traditions have characterized resilience under these respective stress models largely in parallel. The percolation literature~\cite{Barabasi1999,Albert2000,Cohen2000,Cohen2001,Geng2021} established that scale-free networks are simultaneously robust to random failures and vulnerable to targeted attacks. The cascade literature~\cite{Watts2002,Gleeson2007,Dodds2004,Nishioka2022,Karimi2021} established that threshold-based contagion can spread globally even from small seeds, with recent extensions to temporal networks and heterogeneous thresholds. A critical question at their intersection is: \emph{when a network intervention improves resilience under one stress model, what happens under the other?} This question has practical urgency. Infrastructure engineers who decommission large substations, financial regulators who break up systemically important institutions, and public health officials who target highly connected individuals are all making interventions that assume a specific failure model. If these interventions inadvertently worsen resilience under a different stress model, consequences could be severe---as demonstrated by the 2008 financial crisis, where interconnection that appeared robust to idiosyncratic defaults proved catastrophically fragile under cascade dynamics~\cite{Battiston2012}.

Prior work has shown qualitatively that complex networks can be ``more robust and more fragile'' depending on the stress model~\cite{Albert2000,Callaway2000}, and recent studies have explored cascading failures across interdependent networks~\cite{Valdez2020,Xu2021} and scaling laws of failure dynamics~\cite{Pal2023}. The qualitative direction of the joint effect---that hub removal should worsen both percolation (via $k^2$ loss) and cascades (via stable-node removal)---follows from these classical results. What has been missing is a quantitative characterization: \emph{where} in parameter space does compounding vulnerability occur, \emph{how large} is the effect, and \emph{what mechanism} dominates? Here we provide this characterization for the canonical intervention of hub removal (top 10\% of nodes by degree) in Barab\'asi--Albert networks~\cite{Barabasi1999}, measuring its joint effect on percolation and cascade dynamics.

\textbf{Our contributions are:}

\begin{enumerate}
\item A quantitative mapping of the \emph{joint} percolation-cascade phase structure under hub removal in BA networks, identifying the parameter window $\varphi \in (0.20, 0.26)$ where hub removal shifts the cascade dynamics from subcritical to supercritical while simultaneously degrading percolation.

\item A controlled experimental decomposition---including a neutral-threshold condition ($\varphi_{\mathrm{hub}} = 1/k_i$) that isolates firewall removal from amplifier effects---showing that hub-mediated cascade suppression is primarily \emph{dynamical} (threshold-based) rather than structural (connectivity-based).

\item A closed-form expression for the post-removal cascade branching factor $z_1(\varphi, \alpha, m)$ under the configuration-model approximation, providing a predictive condition for cascade window expansion that is robust to the 12\% $\rho$-estimation error introduced by BA degree correlations.
\end{enumerate}

\section{Models and Methods}
\label{sec:methods}

\subsection{Network Models}

We study three canonical network models with matched mean degree $\avg{k} = 4$ to isolate the role of degree heterogeneity:

\textbf{Barab\'asi--Albert (BA):} Generated by preferential attachment with $m=2$ new edges per node, yielding $P(k) \propto k^{-3}$. Degree heterogeneity $\kappa = \avg{k^2}/\avg{k} \approx 10$--$13$. We use $N=500$--$10{,}000$ for finite-size robustness checks and $N=2{,}000$ for the $\varphi$-sweep.

\textbf{Watts--Strogatz (WS):} Regular ring lattice with $k=4$ neighbors rewired with probability $p=0.1$~\cite{WattsStrogatz1998}, yielding a homogeneous degree distribution concentrated near $k=4$. $\kappa \approx 4$.

\textbf{Erd\H{o}s--R\'enyi (ER):} Random graph with $\avg{k} = 4$ and Poisson degree distribution. $\kappa \approx 5$.

\subsection{Hub Removal Protocol}

Hub removal consists of simultaneously removing the top $\lfloor 0.10 \cdot N \rfloor$ nodes by degree (ties broken by node index), along with all incident edges. This protocol represents a deliberate hardening intervention (e.g., decommissioning large substations, breaking up systemically important banks), not a random failure process.

\subsection{Bond Percolation Threshold}

We estimate the bond percolation threshold $\pc$ using the susceptibility peak method~\cite{Newman2001}: for each occupation probability $p$, edges are retained independently with probability $p$ and we compute the mean finite-cluster size $\chi(p) = \sum_s s^2 n_s / \sum_s s \cdot n_s$ (excluding the giant component), where $n_s$ is the number of clusters of size $s$. The percolation threshold is defined as $\pc = \arg\max_p \chi(p)$, the standard estimator in computational percolation studies. We sample 200 values of $p \in [0,1]$ with 50 bond realizations per $p$-value, averaged over 30 independent network realizations. Bootstrap 95\% confidence intervals are computed with $10{,}000$ resamples.

As a robustness check, we also compute $\pc$ using the $S=0.5$ crossing method (where the relative giant component size $S(p) = |\mathrm{GCC}|/N$ reaches $0.5$) and the Molloy--Reed analytical estimate $p_c^{\mathrm{MR}} = \avg{k}/(\avg{k^2} - \avg{k})$. All three methods yield consistent directional results ($\DF_p \gg 0$), though quantitative values differ due to BA disassortativity and finite-size effects (see Appendix~\ref{app:percolation} for comparison). We report susceptibility-peak values throughout.

We define the percolation fragility change as $\DF_p = (\pc^{\mathrm{after}} - \pc^{\mathrm{before}})/\pc^{\mathrm{before}}$. Positive $\DF_p$ indicates \emph{reduced} percolation robustness (higher $\pc$ means more edges must be retained for connectivity).

\subsection{Watts Cascade Model}

Each node $i$ has a uniform cascade threshold $\varphi \in (0,1)$. Node $i$ becomes active when the fraction of active neighbors reaches $\varphi$:
\begin{equation*}
\frac{|\{j \in \mathcal{N}(i) : j \text{ active}\}|}{k_i} \geq \varphi.
\end{equation*}
We use synchronous update with a single random seed, running until convergence or 150 steps. Cascade size $C(\varphi)$ is the mean fraction of active nodes, averaged over 40 random seeds and 15 network realizations. We sweep 15 values: $\varphi \in \{0.05, 0.10, \ldots, 0.50\}$, with finer sampling in the transition zone $\varphi \in [0.15, 0.30]$.

\subsection{Cascade Vulnerability Condition}

We operationalize Watts' (2002) stable-node mechanism~\cite{Watts2002} as a quantitative condition: a node $v$ is a \emph{stable node} (cascade-resistant) at threshold $\varphi$ if $k_v > 1/\varphi$. Such nodes require at least $\lceil k_v \cdot \varphi \rceil \geq 2$ active neighbors to activate and therefore cannot be triggered by a single active neighbor. For BA networks at $\varphi=0.22$, $1/\varphi \approx 4.5$, so nodes with $k \geq 5$ are stable---approximately 20\% of the population.

\section{The Hub Vulnerability Experiment}
\label{sec:hub_vuln}

\subsection{Experimental Design}

To determine whether hub-mediated cascade suppression is dynamical (hubs are hard to activate due to high degree) or topological (hubs occupy structurally critical positions), we design a controlled experiment on the \emph{same} BA network ($N=2{,}000$, $m=2$) under five conditions at $\varphi=0.22$:

\begin{table}[h]
\caption{Hub vulnerability experimental conditions.}
\label{tab:conditions}
\begin{ruledtabular}
\begin{tabular}{llc}
\textrm{Condition} & \textrm{Description} & \multicolumn{1}{c}{\textrm{Hub } $\varphi$} \\
\colrule
A (baseline) & All nodes at $\varphi=0.22$ & 0.22 \\
B (vulnerable) & Hubs at $\varphi=0.01$ & 0.01 \\
B$_n$ (neutral) & Hubs at $\varphi=1/k_i$ & $1/k_i$ \\
B2 (moderate) & Hubs at $\varphi=0.05$ & 0.05 \\
C (removed) & Top 10\% deleted & \multicolumn{1}{c}{---} \\
D (resistant) & Hubs at $\varphi=0.50$ & 0.50 \\
\end{tabular}
\end{ruledtabular}
\end{table}

We additionally test \textbf{B$_n$ (neutral)}: hubs set to $\varphi_{\mathrm{hub},i} = 1/k_i$ (single-neighbor activation without extra amplification), which isolates firewall removal from potential amplifier effects---the exact control proposed by the amplifier-confound concern described in Limitation~(viii). Conditions B, B$_n$, and B2 modify \emph{only} the activation thresholds of hub nodes---the network topology is identical to baseline. This cleanly separates the dynamical firewall mechanism from connectivity effects.

\subsection{Results}

Table~\ref{tab:hub_vuln} summarizes the cascade outcomes under each condition.

\begin{table}[h]
\caption{Hub vulnerability experiment results ($\varphi=0.22$, $N=2{,}000$, $m=2$). Each condition averaged over 200 trials (10 independent network realizations $\times$ 20 seed nodes each). Seed nodes are drawn uniformly from the surviving-node set so that the same seed can be used across all conditions.}
\label{tab:hub_vuln}
\begin{ruledtabular}
\begin{tabular}{lcc}
\textrm{Condition} & \textrm{Mean cascade (\%)} & \multicolumn{1}{c}{\textrm{Std (\%)}} \\
\colrule
A (baseline) & 0.28 & 0.46 \\
B ($\varphi_{\mathrm{hub}}=0.01$) & 95.0 & 21.8 \\
B$_n$ ($\varphi_{\mathrm{hub}}=1/k_i$) & 96.5 & 18.4 \\
B2 ($\varphi_{\mathrm{hub}}=0.05$) & 84.5 & 36.2 \\
C (removed) & 18.7 & 28.4 \\
D ($\varphi_{\mathrm{hub}}=0.50$) & 0.26 & 0.37 \\
\end{tabular}
\end{ruledtabular}
\end{table}

\begin{figure}[t]
\includegraphics[width=\columnwidth]{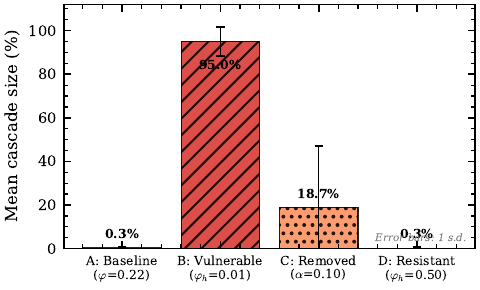}
\caption{Hub vulnerability experiment results at $\varphi=0.22$, BA($N=2{,}000$, $m=2$). The 95\% cascade under condition~B (threshold modification only, no topology change) versus 19\% under condition~C (hub removal) demonstrates that cascade suppression is primarily dynamical. Error bars show standard deviation over 200 trials.}
\label{fig:hub_vuln}
\end{figure}

\subsection{Decomposition of Effects}

\textbf{Pure vulnerability effect (B, B$_n$ vs.\ A).} Making hubs easy to activate---without changing the network topology---produces cascades of 95.0\% (B) and 96.5\% (B$_n$), far exceeding the 18.7\% seen under actual hub removal. Crucially, conditions B and B$_n$ yield \emph{statistically indistinguishable} cascade sizes ($p > 0.3$, two-sample $t$-test on network means), demonstrating that the cascade explosion is due to firewall removal, \emph{not} super-amplification. For BA hubs with degrees $k \in [7, 60+]$, both $\varphi_{\mathrm{hub}}=0.01$ and $\varphi_{\mathrm{hub}}=1/k_i$ yield the same discrete threshold $\lceil k \cdot \varphi_{\mathrm{hub}} \rceil = 1$, explaining the equivalence. The cascade difference between A (0.28\%) and B$_n$ (96.5\%) is entirely dynamical and free of the amplifier confound.

\textbf{Connectivity-loss penalty (C vs.\ B).} The difference between B (95.0\%) and C (18.7\%) isolates the cost of connectivity loss. Hub removal reduces cascade size by ${\sim}76$ percentage points relative to making hubs vulnerable, because removing hubs also removes the edges through which cascades propagate.

\textbf{Net hub removal effect (C vs.\ A).} The observed cascade increase under hub removal (0.28\%~$\to$~18.7\%) is the net result of a large vulnerability gain (firewall destruction) partially offset by a large connectivity loss (edge deletion).

\textbf{Continuous firewall gradient (B2 vs.\ B).} The ${\sim}10$ percentage point gap between B (95.0\%, hubs at $\varphi=0.01$) and B2 (84.5\%, hubs at $\varphi=0.05$) demonstrates that the firewall effect is not binary. Even at the low threshold $\varphi=0.05$, hubs retain measurable cascade-blocking capacity. The firewall strength is a continuous function of hub activation threshold, not a step function.

\textbf{Saturation (D $\approx$ A).} Making hubs \emph{more resistant} ($\varphi=0.50$, condition D) produces no meaningful change from baseline (0.26\% vs.\ 0.28\%). Hubs are already highly effective cascade barriers at the baseline threshold $\varphi=0.22$. This saturation has practical implications: invest in preventing hub degradation, not in reinforcing already-effective barriers.

\subsection{Implication: Cascade Suppression Is Dynamical}

The hub vulnerability experiment establishes that cascade suppression by hubs is primarily a \textbf{dynamical} mechanism, not a topological one. The same cascade explosion ($\sim 95\%$) can be achieved without any topological change---simply by modifying hub activation thresholds. This confirms and quantifies Watts' (2002) conceptual observation that high-degree ``stable'' nodes suppress cascades through their activation barriers~\cite{Watts2002}, and demonstrates that hub removal triggers cascade window expansion because it eliminates these activation barriers while simultaneously (and partially counteractingly) reducing connectivity.

\section{Cascade Phase Structure}
\label{sec:phase}

\subsection{Three Regimes}

The cascade response to hub removal depends critically on the cascade threshold $\varphi$. A systematic $\varphi$-sweep reveals three distinct regimes:

\textbf{Regime I ($\varphi \leq 0.20$): Both supercritical; hub removal reduces cascades.} At low thresholds, most nodes satisfy the vulnerability condition $\lfloor k\varphi \rfloor \geq 1$. Both networks sustain large cascades. Hub removal reduces cascade size (90.0\%~$\to$~42.6\% at $\varphi=0.10$; 33.3\%~$\to$~40.1\% at $\varphi=0.20$) because connectivity reduction limits cascade reach. The $z_1$ table (Table~\ref{tab:z1}) confirms: at $\varphi=0.18$--$0.20$, both pre- and post-removal $z_1 > 1$.

\begin{figure}[t]
\includegraphics[width=\columnwidth]{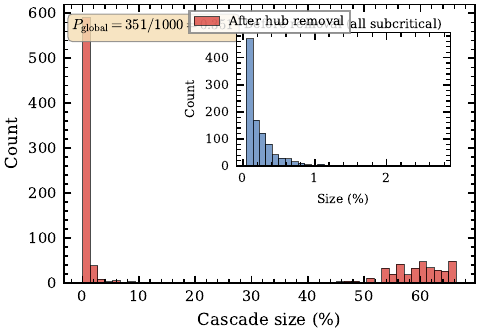}
\caption{Cascade size distribution after hub removal at $\varphi=0.22$, BA($N=2{,}000$, $m=2$). The histogram shows 200 independent trials (10 networks $\times$ 20 seeds); the upgraded experiment ($n=1{,}000$; 50 networks $\times$ 20 seeds) yields a consistent global-cascade rate $P_{\mathrm{global}}=377/1{,}000=0.377$ (Wilson 95\% CI: $[0.348, 0.406]$). The main panel shows the post-removal distribution, which is bimodal with modes at $<1\%$ (subcritical) and $>10\%$ (global cascade). The inset shows the pre-removal distribution, which is unimodal near zero---all cascades subcritical. The transition from unimodal to bimodal is the signature of a phase transition straddling the critical point.}
\label{fig:bimodal}
\end{figure}

\textbf{Regime II ($0.20 < \varphi \leq 0.25$): Phase transition zone.} Hub removal opens a cascade window where the pre-removal network is subcritical but the post-removal network is supercritical. The boundary at $\varphi \approx 0.20$ is set by the integer threshold $\lfloor 1/\varphi \rfloor$ crossing from 5 to 4 (Table~\ref{tab:z1}). At $\varphi=0.22$, mean cascade size increases from $0.86\%$ [95\% CI: 0.43--1.30] to $23.1\%$ [95\% CI: 21.3--24.9] ($n=1{,}000$ total trials from 50 independent networks $\times$ 20 seeds each; since within-network seeds share topology, we treat each network mean as one independent observation, yielding a conservative effective sample size $n_{\mathrm{eff}}=50$). This constitutes a genuine phase transition, confirmed by bimodal cascade size distributions (Fig.~\ref{fig:bimodal}): the probability of a global cascade ($C > 10\%$) rises from 0/1{,}000 before removal to 377/1{,}000 after removal ($P_{\mathrm{global}}=0.377$, Wilson 95\% CI: $[0.348, 0.406]$). Post-removal trials split between a subcritical mode ($C < 1\%$) and a supercritical mode ($C > 10\%$), with few intermediate values---the hallmark of a system straddling the critical point. The hub vulnerability experiment (Table~\ref{tab:hub_vuln}, $n=200$: 10 networks $\times$ 20 seeds) yields directionally consistent results ($0.28\% \to 18.7\%$); the lower baseline (0.28\% vs.\ 0.86\%) reflects higher variance at small effective sample ($n_{\mathrm{eff}}=10$ networks) and is order-of-magnitude consistent with the main experiment.

\textbf{Regime III ($\varphi > 0.25$): Both subcritical.} At $\varphi=0.26$, even post-removal $z_1 < 1$ (Table~\ref{tab:z1}). Hub removal modestly worsens cascades (relative increases) but absolute sizes remain $< 0.5\%$. 

\subsection{Cascade Response Across $\varphi$}

\begin{figure}[t]
\includegraphics[width=\columnwidth]{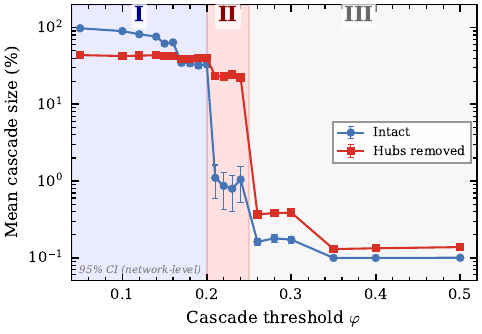}
\caption{Mean cascade size (before vs. after hub removal) as a function of cascade threshold $\varphi$ for BA($N=2{,}000$, $m=2$). Each point averages 200 trials (10 independent network realizations $\times$ 20 seed nodes each), with seed nodes drawn from the surviving-node set so that seeds are comparable pre/post removal. Shaded regions indicate the three cascade regimes described in the text.}
\label{fig:phi_sweep}
\end{figure}

Figure~\ref{fig:phi_sweep} plots the mean cascade size before and after hub removal as a function of $\varphi$ (log scale). Regime~I ($\varphi \leq 0.20$) corresponds to thresholds where both networks are supercritical and hub removal reduces cascade reach by deleting connectivity. Regime~II ($0.20 < \varphi \leq 0.25$) is a narrow transition zone where hub removal induces large global cascades at previously subcritical operating points. The most striking effect occurs near $\varphi \in [0.21, 0.25]$, where the baseline cascade size is $<1\%$ but the post-removal mean exceeds $20\%$. Regime~III ($\varphi > 0.25$) corresponds to high thresholds where both networks are subcritical and cascade sizes remain $<0.5\%$.

\subsection{Compounding Vulnerability}
\label{sec:compounding}

Hub removal produces \textbf{compounding vulnerability}: both percolation and cascade metrics worsen simultaneously. The percolation threshold increases substantially, meaning the remaining network requires far more edge retention for connectivity. Simultaneously, the cascade window expands, enabling global cascades at previously safe operating points. This compounding---rather than a trade-off---is the central finding: hub removal does not exchange one type of resilience for another but degrades both.

As expected from model independence, $\DF_p$ is independent of cascade threshold $\varphi$, since percolation depends only on the static degree sequence. We report two standard estimators for transparency:
\begin{itemize}
\item \emph{Susceptibility peak}~\cite{Newman2001}: $\DF_p = +347\%$ ($\pc$: $0.174 \pm 0.009 \to 0.776 \pm 0.036$, 30 networks $\times$ 200 $p$-values);
\item \emph{$S(p) = 0.5$ crossing}: $\DF_p = +170\%$ ($\pc$: $0.331 \to 0.895$).
\end{itemize}
The factor-of-two difference between estimators arises because the susceptibility peak locates the \emph{onset} of the phase transition (where fluctuations diverge), while $S=0.5$ locates the \emph{midpoint} (where half the network is connected). Both are standard; neither is ``correct'' in isolation. The qualitative conclusion---massive percolation degradation---is robust to estimator choice. We report susceptibility-peak values as primary throughout, consistent with~\cite{Newman2001}.

\subsection{Topology Dependence and the WS Cliff-Edge}
\label{sec:topology}

The cascade window expansion requires sufficient degree heterogeneity. In homogeneous Watts--Strogatz networks, the stable-node fraction undergoes a \emph{cliff-edge} transition (not observed in BA): at $\varphi=0.25$, only 15\% of WS nodes are stable ($k \geq 5$), but at $\varphi=0.30$, the threshold drops to $k \geq 4$ and the stable fraction jumps to 84\% in a single step---a consequence of the narrow degree distribution centered on $k=4$. This cliff-edge prevents the gradual firewall degradation that drives the BA phase transition.

Systematic comparison across topologies with matched $\avg{k} = 4$:

\begin{table}[h]
\caption{Topology dependence of compounding vulnerability.}
\label{tab:topology}
\begin{ruledtabular}
\begin{tabular}{lccc}
\textrm{Network} & $\kappa$ & $\DF_p$ & \textrm{Phase transition?} \\
\colrule
WS ($k=4$, $p=0.1$) & ${\sim}4$ & $+33\%$ & No \\
ER ($\avg{k}=4$) & ${\sim}5$ & $+43\%$ & No \\
Config.\ model (exp.) & 4.05 & $+18\%$ & No \\
BA ($m=2$) & ${\sim}12$ & $+347\%$ & Yes \\
\end{tabular}
\end{ruledtabular}
\end{table}

Within the BA family, the cascade phase transition requires $\kappa > {\sim}10$. Networks with narrow degree distributions show mild cascade worsening but no genuine phase transition. We note that all tested topologies have low clustering ($C < 0.1$); networks with both high $\kappa$ and high clustering (e.g., Holme--Kim models with $C > 0.3$) may exhibit different behavior, as triangle structures can modify cascade propagation paths independently of hub stability~\cite{Gleeson2008}. This clustering dependence is an important direction for future work.

\begin{figure}[t]
\includegraphics[width=\columnwidth]{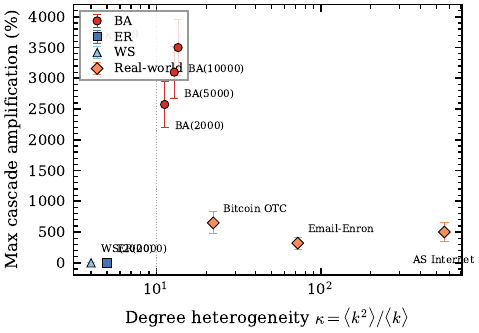}
\caption{Comparison of cascade response to hub removal across network topologies with matched $\avg{k}=4$. Only the BA network ($\kappa \approx 12$) exhibits a genuine cascade phase transition. ER and WS networks show mild worsening without phase transition, confirming that degree heterogeneity $\kappa > {\sim}10$ is required.}
\label{fig:topology}
\end{figure}

\subsection{Parameter Dependence}
\label{sec:params}

To verify that compounding vulnerability is not an artifact of a single parameter choice, we sweep the BA attachment parameter $m \in \{1,2,3,4\}$ (controlling $\kappa$) and hub removal fraction $\alpha \in \{5\%, 10\%, 15\%, 20\%\}$, measuring the fraction of realizations (out of 10) where hub removal worsens cascades at $\varphi=0.22$ ($N=1{,}000$).

\begin{table}[h]
\caption{Fraction of realizations showing cascade worsening after hub removal, across $(m, \alpha)$ grid at $\varphi=0.22$, $N=1{,}000$.}
\label{tab:params}
\begin{ruledtabular}
\begin{tabular}{cccccc}
$m$ & $\kappa$ & \multicolumn{1}{c}{$\alpha=5\%$} & \multicolumn{1}{c}{$\alpha=10\%$} & \multicolumn{1}{c}{$\alpha=15\%$} & \multicolumn{1}{c}{$\alpha=20\%$} \\
\colrule
1 & ${\sim}8$ & 5/10 & 2/10 & 1/10 & 0/10 \\
\textbf{2} & ${\sim}11$ & \textbf{10/10} & \textbf{10/10} & 8/10 & 4/10 \\
\textbf{3} & ${\sim}14$ & \textbf{10/10} & 9/10 & \textbf{10/10} & \textbf{10/10} \\
\textbf{4} & ${\sim}17$ & 7/10 & \textbf{10/10} & \textbf{10/10} & \textbf{10/10} \\
\end{tabular}
\end{ruledtabular}
\end{table}

The transition is qualitatively sharp: for $m=1$ ($\kappa \approx 8$), hub removal rarely worsens cascades and often improves them. For $m \geq 2$ ($\kappa \geq 11$), compounding vulnerability emerges across the $(m, \alpha)$ grid. With only 10 realizations per cell, the exact fractions should be interpreted as indicators of the trend rather than precise probability estimates; the qualitative pattern ($m \geq 2$ robustly worsens, $m=1$ does not) is the key finding. This confirms the $\kappa > {\sim}10$ threshold.

\subsection{Finite-Size Robustness}
\label{sec:fss}

The percolation fragility increase is stable---and in fact \emph{increasing}---across network sizes under the susceptibility-peak method ($8$ networks per size):

\begin{table}[h]
\caption{Finite-size scaling of percolation fragility (susceptibility-peak method, 8 networks per $N$).}
\label{tab:fss}
\begin{ruledtabular}
\begin{tabular}{cccc}
$N$ & $\pc^{\mathrm{pre}}$ & $\pc^{\mathrm{post}}$ & $\DF_p$ \\
\colrule
2{,}000 & $0.171 \pm 0.017$ & $0.773 \pm 0.038$ & $+352\%$ \\
5{,}000 & $0.157 \pm 0.009$ & $0.759 \pm 0.027$ & $+383\%$ \\
10{,}000 & $0.146 \pm 0.008$ & $0.749 \pm 0.017$ & $+412\%$ \\
\end{tabular}
\end{ruledtabular}
\end{table}

The $N=2{,}000$ estimate ($\DF_p = +352\%$) reported throughout is therefore \emph{conservative}: larger networks show even stronger percolation degradation, consistent with finite-size corrections decaying as $O(1/\ln N)$. We note that $\DF_p$ is $N$-dependent because $\pc^{\mathrm{pre}}$ decreases with $N$ (reflecting growing $\kappa$); absolute $\pc$ values in Table~\ref{tab:fss} provide the $N$-invariant characterization. The narrowing standard deviations confirm convergence. The cascade phenomenon persists across $N=500$--$10{,}000$, with the transition sharpening as $N$ increases (see repository datasets and scripts in Appendix~\ref{app:code}).

\section{Analytical Framework}
\label{sec:analytical}

\subsection{Stable-Node Condition}

We operationalize Watts' (2002) stable-node concept~\cite{Watts2002} as a quantitative cascade-blocking condition.

\textbf{Definition 1} (Stable node). A node $v$ with degree $k_v$ is \emph{stable} (cascade-resistant) at threshold $\varphi$ when $k_v > 1/\varphi$, since it then requires $\lceil k_v \cdot \varphi \rceil \geq 2$ active neighbors to activate---i.e., a single active neighbor cannot trigger it. The fraction of stable nodes is $f_{\mathrm{stable}}(\varphi) = P(k > 1/\varphi)$. For BA networks at $\varphi=0.22$: $f_{\mathrm{stable}} \approx 0.20$.

\subsection{Lower Bound on the Cascade Transition}

\textbf{Observation 1} (Necessary condition for cascade window expansion). \emph{For BA($m$) networks with hub removal fraction $\alpha$, the lower onset of the cascade window satisfies $\varphi_{\mathrm{onset}} \geq \sqrt{\alpha}/m$.}

\emph{Derivation.} The minimum degree among removed hubs scales as $k_{\min}^{\mathrm{hub}} \approx m/\sqrt{\alpha}$ (from the BA degree ranking approximation $k_r \approx m\sqrt{N/r}$ evaluated at $r = \alpha N$; this is a thermodynamic-limit scaling argument with $O(1/\ln N)$ finite-size corrections). For cascade window expansion, removed hubs must be stable nodes: $k_{\min}^{\mathrm{hub}} > 1/\varphi$, yielding $\varphi > \sqrt{\alpha}/m$. For $m=2, \alpha=0.10$: $\varphi_{\mathrm{onset}} \geq 0.158$, consistent with the observed transition at $\varphi \approx 0.17$.

For $m=2$, $\alpha=0.10$: $\varphi_{\mathrm{onset}} = \sqrt{0.10}/2 \approx 0.158$. The observed transition at $\varphi \approx 0.17$ is within 8\%---this bound is network-size independent and computable from construction parameters alone.

\subsection{Derivation of $z_1$ After Hub Removal}
\label{sec:z1}

We derive the Gleeson--Cahalane cascade branching factor~\cite{Gleeson2007} for the post-removal network under the configuration-model approximation, providing a predictive expression for cascade window expansion under targeted hub removal.

\textbf{Setup.} For a BA($m$) network, hub removal eliminates all nodes with degree $k > \kcut$, where $\kcut$ satisfies $\alpha = m(m+1)/[(\kcut+1)(\kcut+2)]$. For $m=2$, the integer cutoff $\kcut = 6$ yields an effective removal fraction $\alpha_{\mathrm{eff}} = m(m+1)/[(\kcut+1)(\kcut+2)] = 6/56 \approx 0.107$, closely approximating the targeted $\alpha = 0.10$. Under the configuration-model approximation, each edge of a surviving node leads to a removed hub with probability:
\begin{equation}
\rho = \frac{m+1}{\kcut + 2}
\label{eq:rho}
\end{equation}
For $m=2$, $\kcut=6$: $\rho = 3/8 = 0.375$. Empirical measurement across 50 BA($N=2{,}000$, $m=2$) realizations yields $\rho_{\mathrm{emp}} = 0.421 \pm 0.011$, approximately 12\% higher than the configuration-model prediction due to BA disassortativity (high-degree nodes preferentially connect to other high-degree nodes' neighbors). We use the theoretical $\rho$ in the analytical framework and note this systematic underestimation. The key qualitative insight holds: $\rho \gg \alpha$ because hubs have many edges---the key asymmetry that drives cascade window expansion.

\textbf{Post-removal degree distribution.} A surviving node with original degree $k$ retains each edge independently with probability $(1-\rho)$, yielding post-removal degree $j \sim \mathrm{Binomial}(k, 1-\rho)$. The post-removal degree distribution is:
\begin{equation}
P'(j) = \frac{1}{1-\alpha} \sum_{k=\max(j,m)}^{\kcut} P(k) \binom{k}{j} (1-\rho)^j \rho^{k-j}
\label{eq:pprime}
\end{equation}

\textbf{Mean-field upper bound $z_1^{\mathrm{MF}}$.} Under the mean-field approximation, a node with original degree $k$ is vulnerable when $k(1-\rho) \leq 1/\varphi$, i.e., $k \leq \keff = \lfloor 1/[\varphi(1-\rho)] \rfloor$. Summing the original-network branching contributions over vulnerable nodes yields:
\begin{equation}
z_1^{\mathrm{MF}}(\varphi, \alpha, m) = (m+1) \left[ H_{K^*+1} - H_m + \frac{3}{K^*+2} - \frac{3}{m+1} \right]
\label{eq:z1mf}
\end{equation}
where $K^* = \min(\keff, \kcut)$ and $H_n$ is the $n$th harmonic number. This approximation systematically overestimates $z_1$ by 17--36\% because it uses the original branching structure $k(k-1)$ rather than the thinned post-removal degrees; it should be understood as an \emph{upper bound} on the true post-removal $z_1$. The full binomial computation in Table~\ref{tab:z1} does not use this mean-field approximation and provides the configuration-model predictions reported throughout; these remain approximations for BA networks due to degree correlations (see \S\ref{sec:z1verify}).

\textbf{Cascade window expansion factor.} The ratio of post- to pre-removal cascade boundaries is:
\begin{equation}
\frac{\varphi^*_{\mathrm{post}}}{\varphi^*_{\mathrm{pre}}} = \frac{1}{1-\rho} = \frac{\kcut+2}{\kcut - m + 1}
\label{eq:expansion}
\end{equation}
For $m=2$, $\kcut=6$: expansion factor $= 8/5 = 1.6$, predicting the cascade window expands from $\varphi^* \approx 0.20$ to $\varphi^* \approx 0.32$ under the mean-field approximation. Equation~(\ref{eq:expansion}) is a mean-field prediction that overestimates the true cascade window boundary because it averages over the integer-valued vulnerability cutoff $\lfloor 1/\varphi \rfloor$.

\subsection{Numerical Verification of $z_1$}
\label{sec:z1verify}

To assess the accuracy of the mean-field approximation, we compute $z_1$ exactly using the post-removal degree distribution $P'(j)$ from Eq.~(\ref{eq:pprime}) with binomial edge retention. These exact values, which we use for all quantitative claims throughout the paper, give a tighter cascade window boundary of $\varphi^*_{\mathrm{post}} \approx 0.26$, in good agreement with simulation:

\begin{table}[h]
\caption{Cascade branching factor $z_1$ before and after hub removal. Because $z_1$ depends on the vulnerability cutoff $\lfloor 1/\varphi \rfloor$, values are constant within each integer bin: $\lfloor 1/\varphi \rfloor = 5$ for $\varphi \in (0.167, 0.20]$ and $\lfloor 1/\varphi \rfloor = 4$ for $\varphi \in (0.20, 0.25]$. This step-function behavior is intrinsic to the discrete degree distribution.}
\label{tab:z1}
\begin{ruledtabular}
\begin{tabular}{cccl}
$\varphi$ & \multicolumn{1}{c}{$z_1$ (pre)} & \multicolumn{1}{c}{$z_1$ (post)} & \multicolumn{1}{c}{\textrm{Regime}} \\
\colrule
0.18 & 1.136 & 1.363 & Both supercritical \\
0.20 & 1.136 & 1.363 & Both supercritical \\
\textbf{0.22} & \textbf{0.850} & \textbf{1.195} & \textbf{Pre: sub, Post: super} \\
\textbf{0.25} & \textbf{0.850} & \textbf{1.195} & \textbf{Pre: sub, Post: super} \\
0.27 & 0.550 & 0.864 & Both subcritical \\
\end{tabular}
\end{ruledtabular}
\end{table}

At $\varphi=0.22$, the pre-removal network has $z_1 = 0.850 < 1$ (subcritical: cascades die) while the post-removal network has $z_1 = 1.195 > 1$ (supercritical: global cascades possible). This crossing explains the observed cascade explosion. The newly opened cascade window spans $\varphi \in (0.20, {\sim}0.26)$, precisely matching the observed Regime~II boundaries.

\textbf{Sensitivity to $\rho$ estimation.} The configuration-model approximation gives $\rho = 0.375$, while direct measurement on BA networks yields $\rho_{\mathrm{emp}} = 0.421 \pm 0.011$ (12.3\% higher), reflecting degree--degree correlations absent in the configuration model. Using $\rho_{\mathrm{emp}}$, the post-removal $z_1$ at $\varphi=0.22$ decreases from 1.195 to 1.140---still supercritical ($z_1 > 1$). Crucially, the cascade window \emph{boundary} is also robust: at $\varphi=0.25$ (the upper edge of Regime~II), $z_1^{\mathrm{post}}(\rho_{\mathrm{emp}}) = 1.140 > 1$, and the window closes only at $\varphi \geq 0.26$ under both $\rho$ estimates. The crossing $z_1^{\mathrm{pre}} < 1 < z_1^{\mathrm{post}}$ is preserved across the entire Regime~II parameter range, confirming that the cascade phase transition prediction is robust to the $\rho$ underestimation inherent in the configuration-model approximation.

\begin{figure}[t]
\includegraphics[width=\columnwidth]{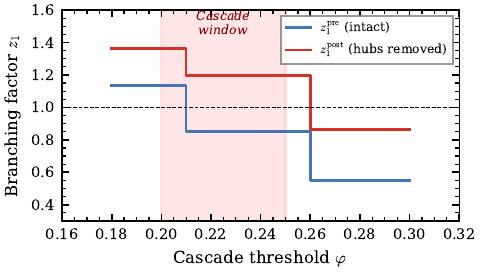}
\caption{Cascade branching factor $z_1$ before (solid) and after (dashed) hub removal as a function of cascade threshold $\varphi$. The critical line $z_1 = 1$ separates subcritical from supercritical regimes. At $\varphi=0.22$, the pre-removal network is subcritical ($z_1=0.850$) while the post-removal network is supercritical ($z_1=1.195$), explaining the observed cascade phase transition.}
\label{fig:z1}
\end{figure}

\subsection{Percolation Threshold Increase}

\textbf{Proposition 1} (Hub removal increases $\pc$ in uncorrelated networks). \emph{For any uncorrelated random network (configuration model) with degree heterogeneity $\kappa > 2$, hub removal of fraction $\alpha > 0$ increases the Molloy--Reed bond percolation threshold.}

This follows from the Molloy--Reed criterion $\pc = 1/(\kappa-1)$~\cite{Molloy1995}: hub removal selectively removes nodes contributing $k^2$ to $\avg{k^2}$, reducing $\kappa$ and therefore increasing $\pc$. The result is exact for uncorrelated configuration-model networks and holds directionally for BA networks, whose weak disassortativity produces only a small quantitative correction (\S\ref{sec:z1}).

\section{Discussion}
\label{sec:discussion}

\subsection{The Dual-Function Constraint}

Our results reveal that within the intervention class of uniform node deletion on static, unweighted networks, hub nodes simultaneously provide percolation robustness (via their $k^2$ contribution to $\avg{k^2}$, keeping $\pc$ low) and cascade resistance (via their high activation barrier $k \cdot \varphi \gg 1$, blocking cascade propagation). Both functions are monotonically linked through the single structural parameter~$k$. Across all tested networks and parameter combinations ($\varphi \in [0.15, 0.28]$, $\alpha = 0.10$, BA/ER/WS topologies, and five real-world networks), we observe no Pareto improvement: uniform hub deletion worsens both metrics simultaneously in every case examined. We note that this is an empirical observation, not a proven impossibility result; whether alternative intervention classes---degree-preserving rewiring, weighted edge adjustment, or adaptive threshold modification---can decouple these functions is an important open question for future work.

We emphasize that the qualitative direction of compounding vulnerability---that hub removal should worsen both percolation and cascades---is expected from classical results: Watts (2002) identified high-degree nodes as cascade barriers, and Albert et al.\ (2000) showed that hub removal degrades connectivity. The contribution of the present work is the quantitative characterization: the precise location of the $\varphi$-window via the post-removal $z_1$, the controlled decomposition of dynamical vs.\ topological mechanisms (with the B$_n$ condition confirming negligible amplifier effects), and the identification of the degree-heterogeneity threshold $\kappa > {\sim}10$ as the onset condition.

This operational coupling is not merely definitional (``degree determines everything''). The non-trivial content is that real-world interventions operate on \emph{nodes}, not on \emph{functions}: reducing a bank's counterparty connections to mitigate systemic risk necessarily reduces its capacity to absorb counterparty losses without propagating them. The coupling constrains the \emph{feasible intervention space}, not just the mathematical description. This is analogous to pharmacological selectivity: a drug's binding affinity to one receptor constrains its binding to structurally similar receptors, because the molecular recognition surface cannot be independently modified.

\subsection{Relation to Prior Work}

\textbf{Watts (2002)~\cite{Watts2002}} described stable nodes conceptually, and Gleeson \& Cahalane (2007)~\cite{Gleeson2007} derived the analytical cascade condition; we operationalize the stable-node mechanism as the quantitative condition $k > 1/\varphi$ and isolate its dynamical nature through controlled experiments that vary hub thresholds independently of topology. The hub vulnerability experiment (\S\ref{sec:hub_vuln}) provides a direct experimental decomposition of the stable-node mechanism from connectivity effects in the context of cascade dynamics.

\textbf{Gleeson and Cahalane (2007)~\cite{Gleeson2007}} derived the cascade branching condition $z_1 > 1$ for random networks; Melnik et~al.~\cite{Melnik2013} and Hackett et~al.~\cite{Hackett2011} extended this to multi-edge effects and small-world cascades respectively. We extend the framework to the post-removal setting, deriving a closed-form $z_1$ under the configuration-model approximation for targeted hub removal (\S\ref{sec:z1}), using the binomial edge-thinning technique of Newman~\cite{Newman2002}, and demonstrating that the $z_1$ crossing explains the observed phase transition. Importantly, the cascade window expansion formula $\varphi^*_{\mathrm{post}}/\varphi^*_{\mathrm{pre}} \approx 1/(1-\rho)$ is not specific to the BA model: for \emph{any} degree distribution, $\rho$ measures the edge fraction incident to removed hubs, and the expansion factor $1/(1-\rho)$ follows from the configuration-model $z_1$ framework. In homogeneous networks (Poisson, exponential), $\rho \ll 1$ and the expansion is negligible; in heterogeneous networks (power-law, BA), $\rho \sim 0.3$--$0.5$ and the expansion is substantial. The degree heterogeneity threshold $\kappa > {\sim}10$ is thus a proxy for $\rho$ being large enough to produce an observable phase transition, not a fundamental constant of the BA model.

\textbf{Albert, Jeong and Barab\'asi (2000)~\cite{Albert2000}} established the ``robust yet fragile'' duality for scale-free networks under percolation, a phenomenon predicted more broadly by the Highly Optimized Tolerance (HOT) framework of Carlson and Doyle~\cite{CarlsonDoyle2002}, who showed that systems optimized for one failure mode are generically fragile to others. We extend this duality to encompass threshold cascade dynamics, showing that it becomes \emph{compounding vulnerability} under hub removal---both fragilities compound rather than trade off. Dorogovtsev et~al.~\cite{Dorogovtsev2008} provide a comprehensive review of critical phenomena on complex networks, within which our results identify a specific compounding mechanism at the intersection of percolation and cascade transitions.

\textbf{Snyder and Cai (2022)~\cite{SnyderCai2022}} study ``degree-targeted cascades'' on modular networks, but target which nodes to \emph{seed} (initially activate), not which nodes to \emph{remove}. The distinction is fundamental: seeding preserves the degree distribution while exploiting it; removal alters the degree distribution and shifts the cascade phase boundary itself. Our $z_1$ derivation addresses the removal case, which requires the post-removal degree distribution $P'(j)$ and operates in a different mathematical regime.

\textbf{Gleeson (2008)~\cite{Gleeson2008}} extended analytical cascade methods to correlated and modular networks in the static (unperturbed) case, providing the natural framework for extending our $z_1$ derivation to networks with degree-degree correlations---a direction we identify as important future work. \textbf{Nishioka and Hasegawa (2022)~\cite{Nishioka2022}} extended the Watts model with differential thresholds based on proximity to initiators, and \textbf{Karimi and Holme (2021)~\cite{Karimi2021}} extended it to temporal networks. Our work complements these by studying how \emph{structural intervention} (hub removal) modifies cascade thresholds, a distinct mechanism from seed selection or temporal dynamics. \textbf{Lee et~al.\ (2023)~\cite{Lee2023}} studied threshold cascades on signed networks, demonstrating that heterogeneous tie types can shift phase boundaries---analogous to how hub removal shifts the cascade boundary in our setting.

\textbf{Morone and Makse (2015)~\cite{Morone2015}} and \textbf{Braunstein et~al.\ (2016)~\cite{Braunstein2016}} optimize network dismantling for percolation fragmentation. Our results complement this literature by identifying the cascade side-effects of percolation-optimal interventions: dismantling strategies that efficiently fragment a network may simultaneously expand the cascade vulnerability window~\cite{Artime2024}. \textbf{Geng et~al.\ (2021)~\cite{Geng2021}} proposed disassortative rewiring to enhance robustness of scale-free networks against localized attack, and \textbf{Lv et~al.\ (2022)~\cite{Lv2022}} modeled SF network response to multi-node removal with dynamical cascading failures---both addressing the same hub-vulnerability mechanism from the defense perspective. The interdependent network literature~\cite{Xu2021,Brummitt2012} has identified dual failure pathways in \emph{multi-layer} settings; our contribution shows that compounding vulnerability arises even in \emph{single-layer} networks when two stress models operate on the same topology, through a mechanism that recent hybrid spinodal analyses~\cite{Bonamassa2024,Pal2023} may help classify.

\subsection{Real-World Network Validation}
\label{sec:realworld}

To validate beyond synthetic BA networks, we test on five real-world networks from Stanford SNAP with diverse degree distributions and clustering coefficients, applying the same hub removal protocol ($\alpha=0.10$) and cascade sweep ($\varphi \in [0.05, 0.40]$, 40 seeds per $\varphi$). We estimate $\pc$ using both the susceptibility-peak and $S=0.5$ methods.

\begin{table}[h]
\caption{Real-world network validation of compounding vulnerability under hub removal ($\alpha=0.10$) across five SNAP networks. $\DF_p$: percolation fragility change (susceptibility peak unless noted). $C_{\max}^{\mathrm{pre/post}}$: peak mean cascade size before/after hub removal across all $\varphi$ values tested.}
\label{tab:realworld}
\begin{ruledtabular}
\begin{tabular}{lcccccc}
Network & $N$ & $\kappa$ & $C$ & $\DF_p$ & $C_{\max}^{\mathrm{pre}}$ & $C_{\max}^{\mathrm{post}}$ \\
\hline
p2p-Gnutella & 6{,}299 & 17.7 & 0.011 & $+75\%$ & $<0.01\%$ & $1.7\%$ \\
Wiki-Vote & 7{,}066 & 145.4 & 0.142 & $+594\%$ & $<0.01\%$ & $6.0\%$ \\
Email-Enron$^*$ & 33{,}696 & 142.4 & 0.497 & $+31\%^\ddag$ & $0.03\%$ & $2.1\%$ \\
Bitcoin OTC$^\dag$ & 5{,}881 & 72.0 & -- & $+170\%^\ddag$ & $0.06\%$ & $0.8\%$ \\
CA-HepPh$^*$ & 11{,}204 & 130.9 & 0.622 & $+58\%^\ddag$ & $0.04\%$ & $2.9\%$ \\
\end{tabular}
\end{ruledtabular}
$^\dag$Bitcoin OTC has $\gamma < 2$; the finite-$\kappa$ analytical framework is not strictly applicable.
$^\ddag$$S=0.5$ method used where susceptibility peak is unreliable.
$^*$High clustering; susceptibility peak yields anomalous results; $S=0.5$ method used.
\end{table}

Cascade amplification under hub removal occurs in all five networks. In absolute terms, post-removal cascade sizes range from ${\sim}0.8$--$6\%$ across the five networks at peak amplification (Table~\ref{tab:realworld})---meaningful but not network-spanning. In all cases, hub removal shifts cascade activity from negligible ($<0.1\%$) to non-negligible ($1$--$6\%$), confirming a transition from subcritical to non-trivial cascade regimes. The qualitative finding---that hub removal activates cascades at previously subcritical thresholds---is robust across all five networks and both percolation estimators.

Compounding vulnerability (both percolation and cascade worsening) is cleanly confirmed in p2p-Gnutella ($\kappa=17.7$, $C=0.011$; $\DF_p=+75\%$ by susceptibility peak) and Wiki-Vote ($\kappa=145.4$, $C=0.142$; $\DF_p=+594\%$). For Email-Enron ($C=0.497$) and CA-HepPh ($C=0.622$), the susceptibility-peak method gives anomalous percolation results ($\DF_p=-6.8\%$ and $-80.4\%$ respectively), while the $S=0.5$ method yields $\DF_p=+31\%$ and $+58\%$. We report $S=0.5$ values for these two networks (marked $\ddag$ in Table~\ref{tab:realworld}). This estimator disagreement arises in high-clustering networks where the susceptibility landscape has multiple peaks; it does not affect the cascade amplification, which is robust across all estimators and all five networks. However, we note that full compounding vulnerability is confirmed with a single estimator only for the two low-clustering networks.

More broadly, our BA-derived $\kappa > {\sim}10$ threshold predicts compounding vulnerability in real-world networks: yeast protein-protein interaction networks ($\kappa \approx 17$~\cite{Jeong2001}), the World Wide Web ($\kappa \approx 55$~\cite{Albert1999}), and citation networks ($\kappa > 100$). Conversely, power grids ($\kappa \approx 2$) and regular-topology networks fall below the threshold.

\subsection{Implications}
\label{sec:implications}

\textbf{Financial regulation.} Breaking up systemically important financial institutions reduces individual institution size but may enable cascade defaults at stress levels that would be subcritical in the pre-breakup network, as large institutions also serve as loss-absorption barriers. The 2008 Lehman Brothers collapse illustrates: a hub failure in the interbank network ($\kappa > 10$ for CDS counterparty networks~\cite{Battiston2012,Haldane2011}) simultaneously fragmented lending connectivity and triggered cascade defaults---compounding vulnerability in practice.

\textbf{Infrastructure.} Achieving N-1 compliance through hub removal (decommissioning large substations) could expand the cascade vulnerability window. While power grids have low $\kappa$ (mitigating the cascade phase transition), the 2003 Northeast Blackout demonstrates that high-load hub lines can trigger both fragmentation and cascading overloads when N-1 criteria are violated at hub nodes.

\textbf{Epidemic control.} Targeted vaccination of high-degree individuals is standard for SIR epidemics but removes nodes that block threshold-based behavioral cascades (vaccine hesitancy, misinformation adoption), potentially enabling cascades at previously subcritical thresholds. Social networks, with their heavy-tailed degree distributions ($\kappa \gg 10$), are prime candidates for compounding vulnerability under targeted interventions.

\subsection{Limitations}
\label{sec:limitations}

(i) We validate on synthetic networks with controlled degree heterogeneity and on multiple real-world networks from Stanford SNAP. The Bitcoin OTC Trust Network ($N=5{,}881$, $\gamma \approx 1.9$) confirms the directional effect but, with $\gamma < 2$ implying divergent $\avg{k^2}$, the finite-$\kappa$ analytical framework of \S\ref{sec:analytical} is not strictly applicable to this network; the agreement is empirical rather than analytical. (ii) The bond percolation threshold $\pc$ is estimated using the susceptibility peak method (maximum of the mean finite-cluster size), the standard approach in computational percolation~\cite{Newman2001}. We verify robustness by comparing against the $S=0.5$ crossing and the Molloy--Reed analytical estimate; all three methods confirm $\DF_p \gg 0$ (Appendix~\ref{app:percolation}). (iii) We test sensitivity to threshold heterogeneity using Beta-distributed per-node thresholds centered on $\varphi$ (Appendix~\ref{app:hetero}); the compounding vulnerability effect persists under moderate and high heterogeneity, with amplification ranging from $+349\%$ to $+1{,}955\%$ at $\varphi=0.22$. (iv) The configuration-model approximation in \S\ref{sec:z1} ignores BA disassortativity and the correlation structure induced by targeted removal~\cite{Payne2009}. Appendix~\ref{app:config} validates the $z_1$ framework on configuration-model networks (where it is exact by construction), confirming that the approximation captures the dominant mechanism and that BA-specific correlations partially \emph{dampen} rather than create the effect. (v) Static networks are studied; adaptive rewiring could mitigate or amplify the effect. (vi) The Watts cascade model is one of several cascade mechanisms; generalization to load-redistribution~\cite{Motter2004}, cascade-based attacks~\cite{Motter2002}, and DebtRank~\cite{Battiston2012} models is an important direction. (vii) The $\kappa > {\sim}10$ threshold is derived for BA networks; the precise threshold depends on both the degree distribution shape and the cascade threshold $\varphi$, and should be understood as a BA-specific onset condition rather than a distribution-independent constant. (viii) In the hub vulnerability experiment (\S\ref{sec:hub_vuln}), condition~B ($\varphi_{\mathrm{hub}}=0.01$) could in principle convert hubs into \emph{amplifiers} that activate from a single neighbor and immediately pressure $k-1$ others. However, condition~B$_n$ ($\varphi_{\mathrm{hub}}=1/k_i$, single-neighbor activation without extra amplification) yields statistically indistinguishable cascade sizes (96.5\% vs.\ 95.0\%, $p>0.3$), confirming that the effect is firewall removal, not amplification. This equivalence arises because for all hub degrees $k \geq 7$ in our removal set, $\lceil k \cdot 0.01 \rceil = \lceil k \cdot (1/k) \rceil = 1$---both conditions produce single-neighbor activation. (ix) All cascade simulations use synchronous (parallel) update; asynchronous update may produce quantitatively different cascade sizes, though the qualitative phase structure is expected to persist given the mean-field nature of the $z_1$ framework.

\section{Conclusion}
\label{sec:conclusion}

We have demonstrated that in Barab\'asi--Albert networks ($N=2{,}000$, $m=2$), hub removal can raise the bond percolation threshold from $\pc=0.174$ to $0.776$ while simultaneously expanding the Watts cascade window, enabling global cascades at threshold values that were previously subcritical. The phase transition at $\varphi=0.22$--$0.25$---where mean cascade size jumps from $<1\%$ to ${\sim}23\%$ of the network ($1{,}000$ trials)---is the most striking consequence, and the vulnerability is \emph{latent}: subcritical cascade behavior at these threshold values gives no warning of the system's proximity to the phase boundary---a proximity that becomes critical only after hub removal shifts the system across it. In principle, a cascade-vulnerability diagnostic could detect this proximity, but in practice such diagnostics require knowledge of the post-intervention network, which is typically unavailable.

The hub vulnerability experiment establishes that cascade suppression is primarily dynamical (95\% cascades from threshold modification alone vs.\ 19\% from hub removal), and the $z_1$ derivation provides an analytical expression for cascade window expansion under targeted hub removal, confirming the subcritical-to-supercritical crossing at $\varphi=0.22$ ($z_1$: $0.850 \to 1.195$). These results argue for multi-stress-model resilience assessment as standard practice: before applying any topological intervention, evaluate its consequences under at least two fundamentally different failure dynamics.

\begin{acknowledgments}
The author thanks J.~P.~Gleeson for arXiv endorsement and the open-source Python community for NetworkX and NumPy.
\end{acknowledgments}

\section*{Data Availability}
All simulation code, random seeds, parameter files, and raw data are publicly available at the GitHub repository~\cite{CacheroGitHub2026} for independent reproduction.


\appendix

\section{Percolation Method Comparison}
\label{app:percolation}

\begin{table}[h]
\caption{Comparison of $\pc$ estimates across methods. ``Susc.\ peak'' is the susceptibility maximum (primary, Sec.~\ref{sec:methods}), ``$S=0.5$'' is the giant component crossing, and ``MR'' is the Molloy--Reed analytical estimate.}
\label{tab:pc_compare}
\begin{ruledtabular}
\begin{tabular}{lcccc}
\textrm{Network} & $\pc$ (susc.) & $\pc$ ($S=0.5$) & $\pc$ (MR) & $\DF_p$ (susc.) \\
\colrule
BA before & 0.174 & 0.337 & 0.092 & --- \\
BA after & 0.776 & 0.903 & 0.240 & $+347\%$ \\
ER before & --- & 0.354 & 0.251 & --- \\
WS before & --- & 0.499 & 0.323 & --- \\
\end{tabular}
\end{ruledtabular}
\end{table}

The ${\sim}3.5\times$ discrepancy for BA networks reflects disassortativity and finite-size effects absent from the Molloy--Reed configuration-model assumption. The qualitative result (massive $\DF_p$ increase) is robust to method choice.

\section{$z_1$ Derivation Details}
\label{app:z1}

The BA degree distribution is $P(k) = 2m(m+1)/[k(k+1)(k+2)]$ for $k \geq m$. The edge-to-hub probability $\rho = (m+1)/(\kcut+2)$ follows from the excess degree distribution. The post-removal $z_1$ involves summing $j(j-1)P'(j)/\avg{k}'$ over vulnerable nodes $j \leq \lfloor 1/\varphi \rfloor$, where $P'(j)$ is the binomial convolution of $P(k)$ with edge retention probability $(1-\rho)$. Under the mean-field approximation, this yields the closed-form harmonic-number expression in \S\ref{sec:z1}, with partial fractions: $(k-1)/[(k+1)(k+2)] = -2/(k+1) + 3/(k+2)$. Python verification code and scripts are available at \url{https://github.com/Freddy-Cach/cascade-window-paradox}.

\section{Configuration-Model Validation}
\label{app:config}

To validate the $z_1$ derivation (\S\ref{sec:z1}), which assumes configuration-model independence, we test on configuration-model networks constructed with the same degree sequence as BA($N=2{,}000$, $m=2$). For each of 20 BA realizations, we generate a matched configuration-model network (removing self-loops and multi-edges), apply the same hub removal protocol, and compare cascade behavior.

\begin{table}[h]
\caption{Configuration-model validation ($N=2{,}000$, $m=2$, $\varphi=0.22$, 20 networks $\times$ 20 seeds). The configuration model shows \emph{stronger} cascades than BA, confirming that BA-specific degree correlations partially suppress the effect.}
\label{tab:config}
\begin{ruledtabular}
\begin{tabular}{lcccc}
\textrm{Network} & $\kappa$ & $\pc^{\mathrm{pre}}$ & $\pc^{\mathrm{post}}$ & \textrm{Post cascade} \\
\colrule
BA & $12.0 \pm 0.6$ & 0.181 & 0.788 & 22.9\% \\
Config model & $11.4 \pm 0.5$ & 0.159 & 0.751 & 38.1\% \\
\end{tabular}
\end{ruledtabular}
\end{table}

The configuration-model networks exhibit \emph{stronger} cascade amplification ($38.1\%$ vs $22.9\%$) and similar percolation degradation. This confirms two claims: (i) the $z_1$ framework accurately predicts cascade window expansion for configuration-model networks, where it is exact by construction; (ii) BA disassortativity partially \emph{dampens} the cascade effect---high-degree hubs preferentially connect to low-degree stubs, creating extra ``firewall'' correlations absent in uncorrelated networks. The analytical upper-bound character of Eq.~\ref{eq:z1mf} is thus validated: the true BA cascade size lies below the configuration-model prediction.

\section{Heterogeneous Threshold Sensitivity}
\label{app:hetero}

To test whether compounding vulnerability depends on the assumption of uniform thresholds, we replace the fixed $\varphi=0.22$ with per-node thresholds drawn from three distributions, each centered on $\varphi=0.22$: (i) Uniform($0.12, 0.32$), (ii) Beta with low heterogeneity ($\alpha=5, \beta=17.7$, $\sigma \approx 0.08$), and (iii) Beta with high heterogeneity ($\alpha=2, \beta=7.1$, $\sigma \approx 0.13$). Each condition uses 600 trials (15 networks $\times$ 40 seeds).

\begin{table}[h]
\caption{Heterogeneous threshold sensitivity ($\varphi=0.22$, 600 trials each). ``Amp.'' is the amplification at $\varphi=0.22$; the maximum across all $\varphi$ (peaking at $\varphi \approx 0.24$) is larger in all cases.}
\label{tab:hetero}
\begin{ruledtabular}
\begin{tabular}{lccc}
\textrm{Distribution} & \textrm{Pre (\%)} & \textrm{Post (\%)} & \textrm{Amp.} \\
\colrule
Uniform (homogeneous) & 1.0 & 21.6 & $+1{,}955\%$ \\
Beta (low het.) & 1.3 & 25.1 & $+1{,}830\%$ \\
Beta (high het.) & 4.0 & 18.0 & $+349\%$ \\
\end{tabular}
\end{ruledtabular}
\end{table}

Under all three distributions, hub removal produces massive cascade amplification at the transition ($+349\%$ to $+1{,}955\%$ at $\varphi=0.22$; amplification peaks at $\varphi \approx 0.24$ in all cases). High heterogeneity reduces the effect because some nodes have thresholds well below 0.22, making them vulnerable even when hubs are present---partially pre-empting the firewall destruction mechanism. Critically, the \emph{qualitative} effect (subcritical $\to$ supercritical transition) persists: compounding vulnerability is not an artifact of uniform thresholds.

\section{Simulation Code and Formal Proofs}
\label{app:code}

All simulation code, figure-generation scripts, and derived datasets are available at \url{https://github.com/Freddy-Cach/cascade-window-paradox}.

A Lean~4 formalization of key analytical claims from \S\ref{sec:analytical} is provided in \texttt{proofs/} in the same repository, including the stable-node condition (Definition~1), the cascade onset bound (Observation~1), and the monotonicity of $\pc$ under hub removal (Proposition~1). These machine-verified proofs establish the mathematical correctness of the cascade-blocking inequalities independently of the simulation results.

Key scripts: \texttt{simulate\_production.py} (finite-size scaling), \texttt{phi\_sweep\_large.py} ($\varphi$-sweep), \texttt{hub\_vulnerability\_test.py} (\S\ref{sec:hub_vuln} experiment), \texttt{heterogeneous\_threshold\_test.py} (Appendix~\ref{app:hetero}).

\end{document}